\documentclass{article}

\usepackage[preprint]{neurips_2024}

\usepackage[utf8]{inputenc} 
\usepackage[T1]{fontenc}    
\usepackage{url}            
\usepackage{booktabs}       
\usepackage{amsfonts}       
\usepackage{amsmath}
\usepackage{amssymb}
\usepackage{nicefrac}       
\usepackage{microtype}      
\usepackage[dvipsnames]{xcolor}         
\usepackage[colorlinks=true, allcolors=Blue]{hyperref}
 \usepackage{natbib}
\usepackage{algorithm}
\usepackage{algpseudocode}
\usepackage{graphicx}
\usepackage{subcaption}
\usepackage{multirow}
\usepackage{float}
\usepackage{comment}

\usepackage{chemfig}
\setchemfig{atom sep=2em}

\title{Thermodynamic free energy map for\\
the non-oxidative glycolysis pathways}

\author{%
  Adittya Pal \\
  Institut for Matematik og Datalogi, Syddansk Universitet \\
  Campusvej 55, 5230 Odense M, Danmark \\
  \texttt{adpal@imada.sdu.dk}
}

\begin{document}
\maketitle

\begin{abstract}
  Designing reaction pathways that maximize the production of a target compound in a given metabolic network is a fundamental problem in systems biology. In this study, we systematically explore the non-oxidative glycolysis metabolic network, guided by the principle that reactions with negative Gibbs free energy differences are thermodynamically favored. We enumerate alternative pathways that implement the net non-oxidative glycolysis reaction, categorized by their length. Our analysis reveals several alternative thermodynamically favorable pathways beyond those reported in experiments. In addition, we identify molecules within the network, such as 3-hydroxypropionic acid, that may have significant potential for further investigation.
\end{abstract}

\section{\label{sec:introduction} Introduction}

Glycolysis is a catabolic pathway that converts carbohydrates to acetylphosphate while releasing energy, catalyzed by enzymes. The pathway that occurs in most organisms does not achieve theoretically optimal atom efficiency because two out of every six carbon atoms in the carbohydrate are released as carbon dioxide, which is lost to the atmosphere. However, synthetically designed pathways have reached the theoretical limit of 100\% carbon efficiency, where carbon atoms are not lost as carbon dioxide molecules \cite{nog_engineer}. These pathways are called non-oxidative glycolysis pathways because carbon atoms are not fully oxidized to carbon dioxide (where the carbon atom has an oxidation number of $+4$ compared to $+3$ in acetylphosphate).

\begin{figure}
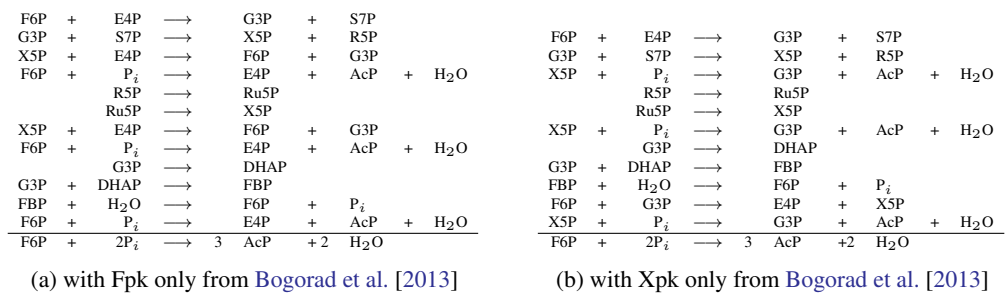
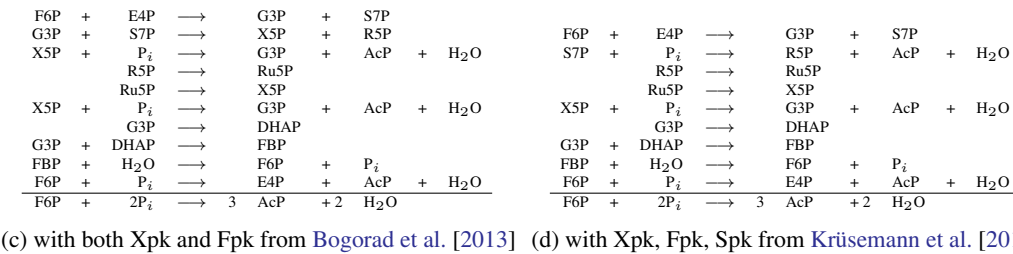

    \centering
    \begin{subfigure}{0.49\textwidth}
    \centering
    \begin{tiny}
        \begin{tabular}{r r r c r l l l l l}
            F6P &+ &E4P &$\longrightarrow$ &&G3P &+ &S7P \\
            G3P &+ &S7P &$\longrightarrow$ &&X5P &+ &R5P \\
            X5P &+ &E4P &$\longrightarrow$ &&F6P &+ &G3P \\
            F6P &+ &P$_i$ &$\longrightarrow$ &&E4P &+ &AcP &+ &H$_2$O \\
            & &R5P &$\longrightarrow$ &&Ru5P \\
            & &Ru5P &$\longrightarrow$ &&X5P \\
            X5P &+ &E4P &$\longrightarrow$ &&F6P &+ &G3P \\
            F6P &+ &P$_i$ &$\longrightarrow$ &&E4P &+ &AcP &+ &H$_2$O \\
            & &G3P &$\longrightarrow$ &&DHAP \\
            G3P &+ &DHAP &$\longrightarrow$ &&FBP \\
            FBP &+ &H$_2$O &$\longrightarrow$ &&F6P &+ &P$_i$ \\
            F6P &+ &P$_i$ &$\longrightarrow$ &&E4P &+ &AcP &+ &H$_2$O \\\hline
            F6P &+ &2P$_i$ &$\longrightarrow$ &3&AcP &+ 2&H$_2$O
        \end{tabular}
    \end{tiny}
    \caption{with Fpk only from \cite{nog_engineer}}
    \label{tab: bogorad_2a}
    \end{subfigure}%
    ~
    \begin{subfigure}{0.49\textwidth}
    \centering
    \begin{tiny}
        \begin{tabular}{r r r c r l l l l l}
            F6P&+&E4P&$\longrightarrow$&&G3P&+&S7P\\
            G3P&+&S7P&$\longrightarrow$&&X5P&+&R5P\\
            X5P&+&P$_i$&$\longrightarrow$&&G3P&+&AcP&+&H$_2$O\\
            &&R5P&$\longrightarrow$&&Ru5P\\
            &&Ru5P&$\longrightarrow$&&X5P\\
            X5P&+&P$_i$&$\longrightarrow$&&G3P&+&AcP&+&H$_2$O\\
            &&G3P&$\longrightarrow$ &&DHAP\\
            G3P&+&DHAP&$\longrightarrow$&&FBP\\
            FBP&+&H$_2$O&$\longrightarrow$&&F6P&+&P$_i$\\
            F6P&+&G3P&$\longrightarrow$&&E4P&+&X5P\\
            X5P&+&P$_i$&$\longrightarrow$&&G3P&+&AcP&+&H$_2$O\\\hline
            F6P&+&2P$_i$&$\longrightarrow$&3&AcP&+2&H$_2$O
        \end{tabular}
    \end{tiny}
    \caption{with Xpk only from \cite{nog_engineer}}
    \label{tab: bogorad_2b}
    \end{subfigure}
    
    \begin{subfigure}{0.49\textwidth}
    \centering
    \begin{tiny}
        \begin{tabular}{r r r c r l l l l l}\\\\\\
            F6P &+ &E4P &$\longrightarrow$ &&G3P &+ &S7P \\
            G3P &+ &S7P &$\longrightarrow$ &&X5P &+ &R5P \\
            X5P &+ &P$_i$ &$\longrightarrow$ &&G3P &+ &AcP &+ &H$_2$O \\
            & &R5P &$\longrightarrow$ &&Ru5P \\
            & &Ru5P &$\longrightarrow$ &&X5P \\
            X5P &+ &P$_i$ &$\longrightarrow$ &&G3P &+ &AcP &+ &H$_2$O \\
            & &G3P &$\longrightarrow$ &&DHAP \\
            G3P &+ &DHAP &$\longrightarrow$ &&FBP \\
            FBP &+ &H$_2$O &$\longrightarrow$ &&F6P &+ &P$_i$ \\
            F6P &+ &P$_i$ &$\longrightarrow$ &&E4P &+ &AcP &+ &H$_2$O \\\hline
            F6P &+ &2P$_i$ &$\longrightarrow$ &3&AcP &+ 2&H$_2$O
        \end{tabular}
    \end{tiny}
    \caption{with both Xpk and Fpk from \cite{nog_engineer}}
    \label{tab: bogorad_2c}
    \end{subfigure}%
    ~
    \begin{subfigure}{0.49\textwidth}
    \centering
    \begin{tiny}
        \begin{tabular}{r r r c r l l l l l}\\\\\\\\
            F6P &+ &E4P &$\longrightarrow$ &&G3P &+ &S7P \\
            S7P &+ &P$_i$ &$\longrightarrow$ &&R5P &+ &AcP &+ &H$_2$O \\
            & &R5P &$\longrightarrow$ &&Ru5P \\
            & &Ru5P &$\longrightarrow$ &&X5P \\
            X5P &+ &P$_i$ &$\longrightarrow$ &&G3P &+ &AcP &+ &H$_2$O \\
            & &G3P &$\longrightarrow$ &&DHAP \\
            G3P &+ &DHAP &$\longrightarrow$ &&FBP \\
            FBP &+ &H$_2$O &$\longrightarrow$ &&F6P &+ &P$_i$ \\
            F6P &+ &P$_i$ &$\longrightarrow$ &&E4P &+ &AcP &+ &H$_2$O \\\hline
            F6P &+ &2P$_i$ &$\longrightarrow$ &3&AcP &+ 2&H$_2$O
        \end{tabular}
    \end{tiny}
    \caption{with Xpk, Fpk, Spk from \cite{nog_engineer2}}
    \label{tab: krusemann}
    \end{subfigure}
    \caption{The sequence of reactions for the non-oxidative glycolytic pathway reported previously with optimal carbon efficiency.}
\end{figure}

While elaborating on the non-oxidative glycolysis pathways, the authors in \cite{nog_engineer} point out that the reaction catalyzed by the enzyme phosphoketolase is an irreversible step, providing the driving force for the non-oxidative glycolytic pathway. The enzyme phosphoketolase can catalyze three different reactions (all following the same template):
\begin{align}
    Xpk: X5P + P_i &\longrightarrow G3P + AcP + H_2O \label{xpk}\\
    Fpk: F6P + P_i &\longrightarrow E4P + AcP + H_2O \label{fpk}\\
    Spk: S7P + P_i &\longrightarrow R5P + AcP + H_2O \label{spk}
\end{align} 
This promiscuous activity of the phosphoketolase enzyme allows for variations in the non-oxidative glycolytic pathway. Three experimentally observed pathways are enumerated in \ref{tab: bogorad_2a}, \ref{tab: bogorad_2b}, and \ref{tab: bogorad_2c}. The fourth pathway, described in \ref{tab: krusemann}, uses the phosphoketolase enzyme to cleave three sugars: xylulose 5-phosphate, fructose 6-phosphate and sedoheptulose 7-phosphate -in a single reaction sequence. \cite{nog_engineer} notes that similar pathways can be derived using sedoheptulose biphosphate instead of fructose biphosphate, combinations of which can generate a large number of pathways for non-oxidative glycolysis. An example of a pathway using sedoheptulose biphosphate was reported in \cite{nog_model_implemen2}.

\setlength{\tabcolsep}{4pt}

We aim to evaluate the possible pathways in the non-oxidative glycolysis metabolic network to develop a physical justification for why some pathways are easier to realize under laboratory conditions than others. Furthermore, our objective is to explain why certain by-products, such as 3-hydroxypropionate in \cite{nog_model_implement}, are observed along with acetylphosphate in experiments attempting to optimize the non-oxidative glycolytic pathway. Finally, we seek to identify potential bottlenecks in these pathways and propose a few that may be the most feasible under practical conditions.

\section{\label{sec:methods} Methods}

Traditionally, constraint-based models have been used to analyze metabolic networks under the steady-state assumption. These constraints-based models can also incorporate kinetic parameters for each (enzymatic) reaction as a further refinement \cite{kinetic_enzyme3,kinetic_enzyme4}. However, these kinetic parameters are often not available for all reactions in the network considered \cite{kinetic_enzyme,kinetic_enzyme2}. Conservation laws, such as conservation of mass, can easily be modeled to ensure that the rate of production of molecules by certain reactions is equal to the rate at which they are consumed in the system. In addition to this conservation law, thermodynamics is another governing principle for most physical processes in equilibrium. An important thermodynamic variable in metabolic networks, in addition to kinetic parameters, is the chemical potential of molecules in the network. For a reaction, $\Delta_r G$ represents the difference in the chemical potentials of the product molecules and the reactant molecules involved in the reaction. Thermodynamics favors only those reactions with a negative $\Delta_r G$ value (often termed reactions that proceed `downhill').

A critical variable affecting the chemical potential of molecules in a system is their concentration. Modeling the concentrations of molecules in living organisms is challenging because they are actively manipulated by transport across membranes through pumps and channels via active transport, which consumes energy in the form of small molecules such as ATP and NADPH. These driven processes are often difficult to assign free-energy differences to, and hence challenging to encapsulate within a network. In addition, different parts of a metabolic network might operate in different cellular compartments with varying concentrations of the same molecule. To avoid these complexities, we assume that the reactions in the metabolic network analyzed occur in a well-stirred bioreactor with the necessary enzymes present, as in a \emph{in vitro} assessment of a hypothesized pathway before it is engineered into living cells. It is admitted that under these assumptions--- not accounting for the free-energy differences of the transport processes and the differential concentration of the same molecule in different cellular compartments--- we would introduce errors into the modeling.

\subsection{Constructing the Reaction Network}

\begin{algorithm}[h!]
    \caption{Generating the Reaction Network}
    \begin{algorithmic}[1]
        \State Define $P_0:$ source collection of molecules as graphs
        \State Define $\mathcal{R}:$ set reaction templates as graph transformation rules $X \overset{r}\longrightarrow Y$, where $r\in \mathcal{R}$ transforms subgraph $X$ to subgraph $Y$
        \State Define constraints as right predicates, $c_i: p\rightarrow\{0, 1\}$, where constraint $c_i$ decides if graph $p$ (with subgraph $Y$) is allowed or not.
        \State i = 0 \Comment{recursion step counter}
        \While{graph $p\in P_i$ satisfies constraint $c_i$}
            \State i++ \Comment{increment step}
            \State $P_i = \left\{q\in Q: p\overset{r}\rightarrow q, p\subset \bigcup_{0\leq j < i} P_j, r\in \mathcal{R}\right\}$ 
            \Statex (this expands the set of existing molecules $P_i$ at the recursion step $i$, by producing a collection of new molecules $Q$ through the application of the rules $r$) 
        \EndWhile
        \State return $\bigcup_{0\leq i} P_i$ \Comment{union of sets of molecules from all steps}
    \end{algorithmic}
    \label{generate_network}
\end{algorithm}

The metabolic reaction networks are generated using the graph grammar approach in M{\O}D, as elaborated in \cite{rule_composition_mod, rule_composition_mod2}. The generic expansion of the molecular space was achieved through the recursive application of reaction templates, represented as graph transformation rules, on molecules represented as graphs. The procedure is illustrated in Algorithm \ref{generate_network} and is detailed in \cite{mod}. The vertices in these (molecular) graphs correspond to atoms, and the edges represent bonds. Hence, subraphs would correspond to molecular fragments. However, this approach omits stereoinformation for the molecules, which is often critical for metabolic reactions. As a result, stereoisomeric carbohydrates with identical graph representations, such as erythrose and threose, ribose and xylose, or ribulose and xylulose, cannot be distinguished. The graph transformation rules that model each enzymatic reaction are listed in Table \ref{tab: list_of_rules} and elaborated in the Appendix \ref{secA1}. The molecular space was expanded from a set of starting molecules listed in Table \ref{tab:mol_table} in the Appendix \ref{secA1}, until no new molecules with fewer than eight carbon atoms could be added. This restriction makes the molecular space finite and reasonable, as carbohydrates with eight or more carbons are typically unstable and rarely observed in metabolic pathways. The expanded molecular space contains $81$ molecules, and the reaction networks generated consisted of $414$ individual reactions.

\begin{table}[t]
    \centering
    \begin{tabular}{r | r  l}
        Abbrev. & Name of the Enzyme & Function  \\\hline
        AL & Aldolase & performs a generic aldol addition. \\
        AlKe & Aldose-Ketose & isomerizes aldehydes to ketones. \\
        KeAl & Ketose-Aldose & isomerises ketones to aldehydes. \\
        PHL & Phosphohydrolase & hydrolyses the phosphate group. \\
        PK & Phosphoketolase & breaks C-C keto bond, adds phosphate.\\
        TAL & Transaldolase & breaks an aldol bond (generally C3).\\
        TKL & Transketolase & cleaves a keto group (generally C2).\\
    \end{tabular}
    \caption{Graph grammar transformation rules for expanding the molecular space.}
    \label{tab: list_of_rules}
\end{table}

\subsection{Enumerating the relevant pathways}

Pathways in living organisms metabolize carbohydrates such that two carbon atoms of a six-carbon sugar (e.g. fructose) are lost as carbon dioxide molecules. Our goal is to identify pathways that do not produce carbon dioxide from the initial starting carbohydrates but instead convert the six carbon atoms of the carbohydrate molecule into three acetylphosphate molecules. Thus, we set a search query that restricts the inflow of fructose 6-phosphate to one and the outflow of acetylphosphate to three. If the number of acetylphosphate molecules produced by the pathway is less than three, the carbon efficiency of the pathway would be suboptimal. The search query is formulated as an integer linear program (ILP) to assign integral flows to the reactions, representing the number of times each reaction must occur in the pathway to produce the target molecules from the given source molecule. The mass conservation law is represented as a flow conservation constraint for all vertices $v$, where the flow $f_e$ on a hyperedge $e$ is the number of times the reaction represented by the hyperedge occurs in the pathway. For all vertices, the number of copies of the molecule used in the reactions $e$ where $v$ is a reactant ($\text{in}(v)$) must be balanced by the number of copies of the molecule produced by the reactions $e$ where $v$ is a product ($\text{out}(v)$) in steady state. Mathematically,    
\begin{equation}
    \forall v\in V: \sum_{e\in \text{in}(v)} f_e = \sum_{e\in \text{out}(v)} f_e
    \label{flow_conservation}
\end{equation}

Multiplying the flow $f_e$ by a positive integer also produces a valid flow that satisfies the mass conservation law. Therefore, inflow is restricted to one fructose 6-phosphate molecule to eliminate superfluous solutions, which are base solutions scaled by a positive integer. An additional constraint ensures that all carbohydrate molecules in the pathway must have a phosphate subgraph, making the returned pathways more realistic. The ILP formulated to search for non-oxidative glycolysis pathways was as follows:

\begin{align}
    \min\left(\sum_{e \in E} f_e \right)& \label{proposed_objective} \\
    \forall e \in E\colon f_e &\geq 0 \label{flow_bounds} \\
    \forall v\in V \colon \sum_{e\in \text{out}(v)} f_e - \sum_{e\in \text{in}(v)} f_e &= 0 \tag{\ref{flow_conservation}}\\
    \text{inFlow}[\text{F6P}] &= 1 \\
    \text{outFlow}[\text{AcP}] &= 3 \label{outflow_constraint}
\end{align}

For the pathways to 3-hydroxypropionic acid, constraint \eqref{outflow_constraint} was replaced by $\text{outFlow}[\text{HPA}] = 3$.
The formulated ILP had $510$ variables, $184$ constraints and was solved using a generic ILP solver; in this case, the open source Coin-OR CBC solver \cite{cbc} was used. The ILP solver took $\sim 190$ seconds to solve for the two pathways using seven reactions described in Section \ref{seven}. The ILP enumerates pathways based on the number of distinct reactions utilized, not the total number of reactions in the pathway. Therefore, if a reaction is used twice (with a flow of two), it is counted only once when ordering the pathways returned by the ILP solver.

\subsection{Assignment of chemical potentials to vertices}

Representing molecules as graphs preserves the connectivity information of the atoms but does not include details about the relative placement of the atoms in the three-dimensional space. To address this, we generated a three-dimensional embedding for the atoms in the molecular graph and used that embedding to assign a chemical potential to the molecule under standard physical conditions. OpenBabel \cite{open_babel} was used to create an initial guess of the three-dimensional embedding of the molecule using a combination of elementary rules and frequently encountered molecular subgraphs. However, these initial assumptions often contain substructures with steric clashes or high strain and, therefore, require optimization.

The torsional angles in the molecule are adjusted to locate a lower energy conformer, where the energy of the resulting conformer is estimated using an elementary force field. In this work, the universal force field (UFF) \cite{uff} was used. The Cartesian coordinates of the optimized embedding of the molecule were written as a \texttt{.sdf} file.

\begin{figure}[h!]
    \centering
    \includegraphics[width=0.8\linewidth]{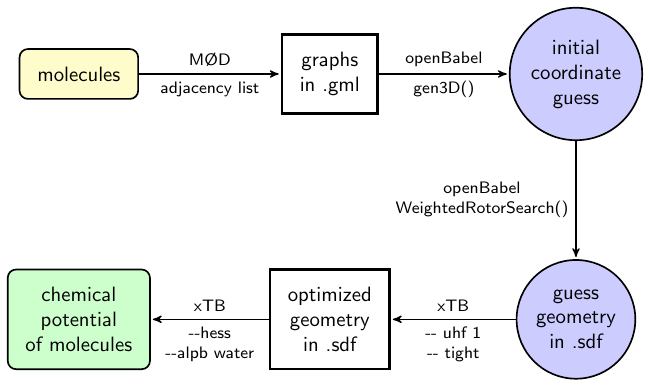}
    \caption{Sequence of operations to estimate the chemical potential of a molecule at standard physical conditions ($273.15$ K and $10^5$ Pa) from the connectivity information of a graph that represents a molecule.}
    \label{fig: thermo_workflow}
\end{figure}

The structures generated by OpenBabel cannot be directly used to estimate the chemical potentials because they are based on empirical knowledge of a class of molecules. Semiempirical methods, such as xTB \cite{grimme_xtb}, offer a fast yet accurate way to estimate thermodynamic parameters after optimizing the geometry of the molecule (with a reasonable initial guess). However, as a chemical tool, xTB requires the input file of the molecule to be in one of the standard chemical formats. OpenBabel serves as a bridge, converting graphs with only connectivity information into standard structure data files readable by xTB. xTB iteratively optimizes the geometry of the molecule from the input \texttt{.sdf} file until the energy change between iterations falls below an acceptable tolerance. The optimized geometry is then used to assign the chemical potential to the molecule. xTB calculates the thermostatistical translational, rotational, and vibrational contributions to the electronic energy to determine the chemical potentials, using the rigid-rotor and harmonic oscillator approximations.

The workflow for assigning chemical potentials to molecules is shown in Figure \ref{fig: thermo_workflow}.

\subsection{Thermodynamic analyses of pathways}

The pathways enumerated by the ILP solver satisfy the constraints of mass conservation, but they must also be realizable in practical reactor-like settings. Although practical realizability (encapsulated as whether a reaction is thermodynamically favorable, i.e., has a negative Gibbs free energy difference under certain concentrations of the involved molecules) can be integrated into the ILP modeling as in \cite{thermoflow,thermoFlow2}, in this work, we perform a post-processing of the ILP solutions using the calculated chemical potentials. We borrow concepts from probability to understand how energy tends to distribute itself in the reaction network, making some reactions thermodynamically favorable. At first glance, this might seem contradictory, as physical laws are universal and deterministic, not probabilistic.

We consider the following statement for the second law of thermodynamics: 
\begin{quote}
    When a system is composed of \emph{large number of molecules}, it is \emph{exceedingly likely} \emph{spontaneously} to move to a thermodynamic macrostate which \emph{maximizes} the number of possible microstates. 
\end{quote}
The caveat is that the system must consist of a large number of molecules, which might not always be the case in cellular compartments in living organisms. Statistics become overwhelming (so they can be formulated as a statistical law of large numbers) only when the number of molecules in the system is large; hence, the justification for using the phrase \emph{exceedingly likely}.

Not all microstates available to the system are equally probable. The probability of a microstate depends on its energy $E$, through the Boltzmann factor $\sim \exp(-E/k_B T)$. Furthermore, there may be several microstates with the same energy, which is related to the entropy $S$ of the system by the Boltzmann-Planck equation: $S = k_B \ln w$, where $w$ is the number of microstates in the system and $k_B$ is the Boltzmann constant. Therefore, the probability $P(E, T)$ that a system is in a macrostate with energy $E$ and $w$ microstates at temperature $T$ is given by the following:
\begin{equation}
    P(E, T) \sim w\cdot\exp(-E/k_B T) = \exp(S/k_B)\cdot\exp(-E/k_B T) = \exp\left(-\frac{E - TS}{k_B T}\right)
\end{equation}
In the context of reactions, we are often concerned with the relative probabilities of two macrostates. The ratio of probabilities of two macrostates with energy $E_1$ and $E_2$ is given by:
\begin{equation*}
    \frac{P(E_1, T)}{P(E_2, T)} = \exp\left(-\frac{(E_1 - E_2) - T (S_1 - S_2)}{k_B T}\right)
\end{equation*}
Under conditions of constant pressure (commonly encountered in cellular and bioreactor environments), the energy $E$ can be replaced by the enthalpy $H$. The numerator of the exponent then becomes $(H_1 - H_2) - T (S_1 - S_2) = \Delta H - T \Delta S = \Delta_r G$. The quantity $\Delta_r G$ is the Gibbs free energy difference for the reaction and indicates the relative likelihood that the reaction will proceed in a particular direction.

For a reaction $r : \text{LHS} \longrightarrow \text{RHS}$, if $\Delta_r G < 0$, then the ratio $P_{\text{RHS}} / P_{\text{LHS}} > 1$, meaning the system is more likely to be in the state represented by the RHS. In contrast, if $\Delta_r G > 0$, the system prefers the state represented by the LHS.

For pathways, the Gibbs free energy is a state function and thus independent of the sequence of reactions. Therefore, all pathways effecting the net reaction for non-oxidative glycolysis:
\begin{equation*}
    \text{frustose 6-phosphate } + 2\text{ phosphoric acid } \longrightarrow 3\text{ acetyl phosphate} + 2H_2O
\end{equation*}
will have the same net Gibbs free energy difference, given by $\Delta_r G = 3G(AcP) + 2G(H_2O) - G(F6P) - 2G(P_i)$. However, different pathways use different intermediate reactions, resulting in varying intermediate compositions and thermodynamic free energy profiles. Ideally, all of the constituent reactions in a pathway should be spontaneous in the required direction. For consecutive reactions with negative Gibbs free energy differences, the molecules on the RHS of the first reaction are more likely to appear, forming a subset that serves as the LHS of the subsequent reaction. This establishes a cascade of reactions in which molecules are consecutively transformed into ones with a high probability of being observed, provided that all reactions in the pathway have negative Gibbs free energy differences.

In contrast, a reaction with a positive Gibbs free-energy difference represents a bottleneck. A reaction $i$ with $\Delta_r G_i > 0$ does not imply that it will \emph{not} occur. This simply means that the reaction is less likely to proceed in the required direction compared to the reverse reaction. In cellular environments, energy-driven processes and transport mechanisms actively manipulating molecule concentrations can render such reactions thermodynamically favorable under those conditions, even if they are not under standard physical conditions. Furthermore, the chemical potentials of the molecules, from which the free-energy differences are calculated, are estimated under certain approximations. Errors in these estimations could lead to incorrect conclusions.

\section{Results}

We enumerate pathways in the generated reaction network, and the distribution of the pathways by the number of reactions used, the number of unique reactions used, and the biphosphate used is shown in Table \ref{tab: pathways_length}.
Intuitively, we hypothesize that pathways utilizing a smaller number of reactions (`shorter paths' in the network) might be favored because it would be simpler to construct a corresponding bioreactor if each reaction takes place in a separate compartment. In addition, pathways that use fewer reactions would be easier to study and monitor under laboratory conditions.

\begin{table}[ht] 
    \centering    
    \begin{tabular}{c | c | c c c | c c c c}
        \multicolumn{2}{c}{Type of biphosphate used} & \hspace{0.2cm}XBP\hspace{0.1cm} & \hspace{0.1cm}FBP\hspace{0.1cm} & \hspace{0.1cm}SBP\hspace{0.2cm} & \multicolumn{4}{c}{Total} \\\hline
        \hspace{0.2cm} 7 reactions \hspace{0.2cm} & \hspace{0.2cm}7 unique reactions\hspace{0.2cm} & 1 & 0 & 1 & \hspace{0.2cm} & & \hspace{0.2cm} & 2 \\\hline
        \multirow{2}{*}{\hspace{0.2cm} 8 reactions \hspace{0.2cm}} & 7 unique reactions & 6 & 0 & 6 & & 12 & & \multirow{2}{*}{29} \\
        & 8 unique reactions & 7 & 2 & 8 & & 17 \\\hline
        \multirow{3}{*}{\hspace{0.2cm} 9 reactions \hspace{0.2cm}} & 7 unique reactions & 3 & 0 & 3 & & 6 & & \multirow{3}{*}{207} \\
        & 8 unique reactions & 42 & 12 & 48 & & 102 & \\
        & 9 unique reactions & 38 & 17 & 44 & & 99 & \\\hline
    \end{tabular}
    \caption{Count of the different pathways enumerated by the ILP solver.}
    \label{tab: pathways_length}
\end{table}

\subsection{\label{seven} Pathways with seven reactions}

\begin{figure}
    \centering
    \begin{subfigure}{\textwidth}
        \includegraphics[width=\linewidth]{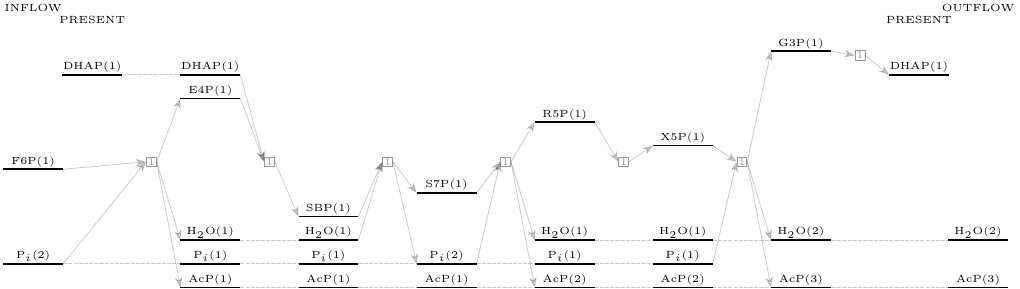}
        \caption{\label{fig: 7reaction_7biphosphate}}
    \end{subfigure}
    \begin{subfigure}{\textwidth}
        \includegraphics[width=\linewidth]{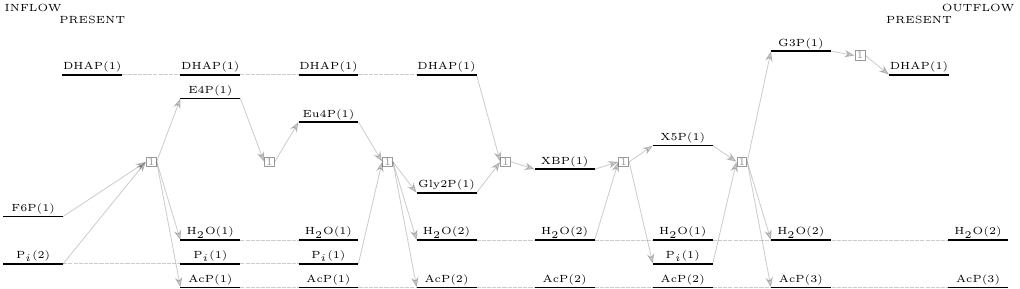}
        \caption{\label{fig: 7reaction_5biphosphate}}
    \end{subfigure}
    \caption{Two pathways for non-oxidative glycolysis using $7$ reactions: \ref{fig: 7reaction_7biphosphate} using sedoheptulose-1,7-biphosphate (top panel) and \ref{fig: 7reaction_5biphosphate} using xylulose-1,5-biphosphate (bottom panel).}
\end{figure}

Two non-oxidative glycolytic pathways with seven reactions were found in the expanded reaction network and are shown in Figure \ref{fig: 7reaction_7biphosphate} and Figure \ref{fig: 7reaction_5biphosphate}. The first uses sedoheptulose-1,7-biphosphate, while the second uses xylulose-1,5-biphosphate. The first pathway has been studied in \cite{nog_model_implemen2} under laboratory conditions and was reported to "overcome bottlenecks from the previously reported non-oxidative glycolysis cycles." Both pathways use only four of the seven templated enzymatic reactions used to expand the reaction network: aldolase, aldose-ketose isomerase, phosphohydrolase, and phosphoketolase.

\begin{figure}[ht]
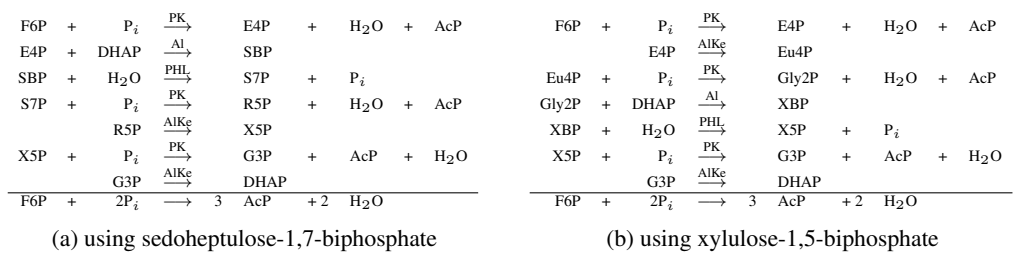

    \centering
    \begin{subfigure}{0.49\textwidth}
    \centering
    \begin{tiny}
        \begin{tabular}{r r r c r l l l l l}
            F6P &+ &P$_i$ &$\overset{\text{PK}}\longrightarrow$ &&E4P &+ &H$_2$O &+ &AcP \\
            E4P &+ &DHAP &$\overset{\text{Al}}\longrightarrow$ &&SBP \\
            SBP &+ &H$_2$O &$\overset{\text{PHL}}\longrightarrow$ &&S7P &+ &P$_i$ \\
            S7P &+ &P$_i$ &$\overset{\text{PK}}\longrightarrow$ &&R5P &+ &H$_2$O &+ &AcP \\
            & & R5P &$\overset{\text{AlKe}}\longrightarrow$ &&X5P \\          
            X5P &+ &P$_i$ &$\overset{\text{PK}}\longrightarrow$ &&G3P &+ &AcP &+ &H$_2$O \\
            & &G3P &$\overset{\text{AlKe}}\longrightarrow$ &&DHAP \\\hline
            F6P &+ &2P$_i$ &$\longrightarrow$ &3&AcP &+ 2&H$_2$O
        \end{tabular}
    \end{tiny}
    \caption{using sedoheptulose-1,7-biphosphate}
    \label{tab: 7react_7BP}
    \end{subfigure}%
    ~
    \begin{subfigure}{0.49\textwidth}
    \centering
    \begin{tiny}
        \begin{tabular}{r r r c r l l l l l}
            F6P &+ &P$_i$ &$\overset{\text{PK}}\longrightarrow$ &&E4P &+ &H$_2$O &+ &AcP \\
            & & E4P &$\overset{\text{AlKe}}\longrightarrow$ &&Eu4P \\
            Eu4P &+ &P$_i$ &$\overset{\text{PK}}\longrightarrow$ &&Gly2P &+ &H$_2$O &+ &AcP \\
            Gly2P &+ &DHAP &$\overset{\text{Al}}\longrightarrow$ &&XBP \\
            XBP &+ &H$_2$O &$\overset{\text{PHL}}\longrightarrow$ &&X5P &+ &P$_i$ \\
            X5P &+ &P$_i$ &$\overset{\text{PK}}\longrightarrow$ &&G3P &+ &AcP &+ &H$_2$O \\
            & &G3P &$\overset{\text{AlKe}}\longrightarrow$ &&DHAP \\\hline
            F6P &+ &2P$_i$ &$\longrightarrow$ &3&AcP &+ 2&H$_2$O
        \end{tabular}
    \end{tiny}
    \caption{using xylulose-1,5-biphosphate}
    \label{tab: 7react_5BP}
    \end{subfigure}
    \caption{The sequence of reactions with the respective enzymes for the non-oxidative glycolytic pathway using $7$ reactions.}
\end{figure}

We elaborate on the individual reactions in the pathways here. Both pathways have two reactions that require one substrate molecule, while the remaining five reactions require two substrate molecules. Only one reaction requires two carbohydrate molecules as a substrate, which has been highlighted in both pathways. All other bimolecular reactions use either water or phosphate molecules, which are present in abundance. Steady-state simulations of the non-oxidative glycolytic pathways by \cite{nog_model_implemen2} have shown that the intermediate sedoheptulose-7-phosphate starts to accumulate over time, limiting the efficiency of the pathway. They addressed this problem by introducing a specific phosphoketolase into the system, which removed the kinetic bottleneck in the pathway. The second pathway with seven reactions, shown in Table \ref{tab: 7react_5BP} found in the reaction network, does not produce sedoheptulose-7-phosphate at all. However, a drawback of this pathway is that although phosphoketolase has been widely reported to be active on substrates such as fructose-6-phosphate, xylulose-5-phosphate \cite{pkp_alternative}, and sedoheptulose-7-phosphate \cite{nog_model_implemen2}, it has not been reported to act on erythrulose-4-phosphate, as required by the third reaction in the pathway shown in \ref{tab: 7react_5BP}. However, if the presence of erythrulose-4-phosphate, glycolaldehyde-2-phosphate and xylulose-1,5-biphosphate is found in a reactor that affects the non-oxidative glycolytic pathway with optimal carbon efficiency using only these four enzymes, it would act as evidence that the pathway shown in \ref{tab: 7react_5BP} is also practical.

\begin{figure}
    \centering
    \begin{subfigure}{0.49\textwidth}
        \includegraphics[width=\linewidth]{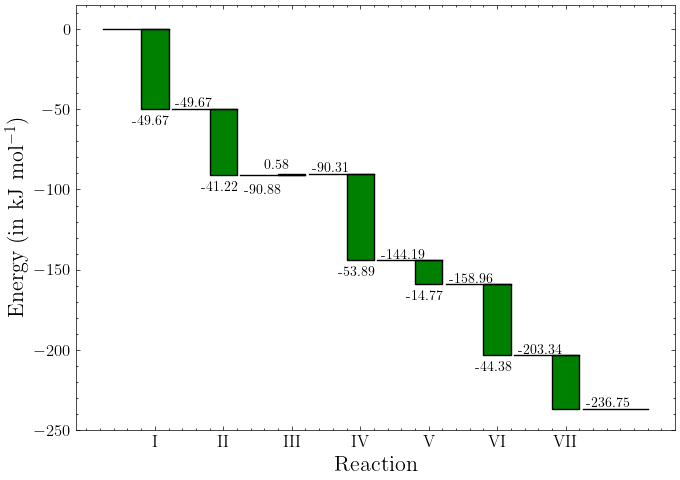}
        \caption{\label{fig: 7reaction_7biphosphate_profile}}
    \end{subfigure}%
    ~
    \begin{subfigure}{0.49\textwidth}
        \includegraphics[width=\linewidth]{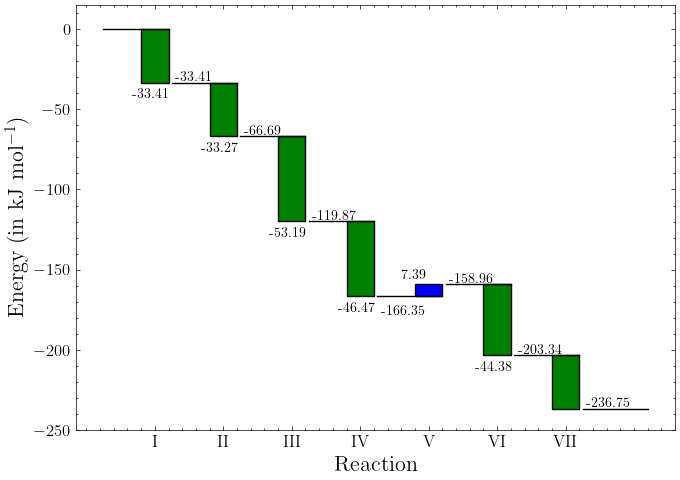}
        \caption{\label{fig: 7reaction_5biphosphate_profile}}
    \end{subfigure}
    \caption{Thermodynamic free energy profiles for the two pathways for non-oxidative glycolysis using $7$ reactions for pathways \ref{tab: 7react_7BP} (left) and \ref{tab: 7react_5BP} (right) respectively.}
\end{figure}

We determined the free energy differences for each reaction used in the two pathways in \ref{tab: 7react_7BP} and \ref{tab: 7react_5BP}. The thermodynamic free energy profiles for these two pathways are plotted in Figures \ref{fig: 7reaction_7biphosphate_profile} and Figure \ref{fig: 7reaction_5biphosphate_profile}, respectively. The green bars represent the Gibbs free energy differences for thermodynamically favorable reactions ($\Delta_r G < 0$), while the blue bars represent the Gibbs free energy differences for reactions that are not thermodynamically favorable ($\Delta_r G > 0$) under standard physical conditions. The bars are annotated with the numerical values of the Gibbs free energy differences for the respective reactions. Since free energy is a state function, the net difference in Gibbs free energy for the overall pathway is the same for both pathways: $-236.75$ kJ mol$^{-1}$.

In both pathways, the hydrolysis of the biphosphate by the phosphohydrolase is not a thermodynamically favorable reaction. Although the hydrolysis of sedoheptulose-1,7-biphosphate has a positive Gibbs free energy difference of $0.58$ kJ mol$^{-1}$, the hydrolysis of xylulose-1,5-biphosphate has a higher value of $7.39$ kJ mol$^{-1}$. The hydrolysis reaction has been observed to be sensitive to the pH of the environment and the activity of water molecules. The hydrolysis of the biphosphate is expected to be a thermodynamic bottleneck for both pathways. In reality, the cell might adjust the physical conditions in the region where the hydrolysis takes place to make the reaction favorable. Moreover, the free energy of the educts and the products is not simply the sum of those of individual molecules because of the complex intermolecular interactions (hydrogen bonding, polar interactions, and hydrophobic effects) that might take place between molecules, especially in solution.

\subsection{\label{experiment} Previously reported pathways}

In this section, we analyze the four previously reported non-oxidative glycolysis pathways listed in Tables \ref{tab: bogorad_2a}, \ref{tab: bogorad_2b}, \ref{tab: bogorad_2c} and \ref{tab: krusemann} and compare them to the shortest pathways described in Section \ref{seven}. Note that the count of reactions in Table \ref{tab: pathways_describe}, listing the various characteristics of the pathways, is one less than that of the pathways listed in Tables \ref{tab: bogorad_2a}, \ref{tab: bogorad_2b}, \ref{tab: bogorad_2c}, and \ref{tab: krusemann} because we cannot represent stereoinformation in molecular graphs. Hence, the molecules ribulose-5-phosphate and xylulose-5-phosphate have the same representation as graphs. The two reactions $R5P \longrightarrow Ru5P \longrightarrow X5P$ are counted as one. All pathways use fructose-1,6-biphosphate.

\begin{table}[!h]
    \centering
    \begin{tabular}{l | c c c c}
        Pathway\hspace{0.5cm} & \hspace{0.5cm}\ref{tab: bogorad_2a}\hspace{0.5cm} & \hspace{0.5cm}\ref{tab: bogorad_2b}\hspace{0.5cm} & \hspace{0.5cm}\ref{tab: bogorad_2c}\hspace{0.5cm} & \hspace{0.5cm}\ref{tab: krusemann}\hspace{0.5cm} \\\hline
        \#reactions\footnotemark[1] & 11 & 10 & 9 & 8 \\
        \#hyperedges\footnotemark[2] & 8 & 8 & 8 & 8 \\
        \#enzymes\footnotemark[3] & 7 & 7 & 7 & 6 (no transketolase) \\
        \#pk type\footnotemark[4] & (0,3,0) & (3,0,0) & (2,1,0) & (1,1,1) \\
        \#bimolecular\footnotemark[5] & 9 & 8 & 7 & 6 \\
        \#two sugars\footnotemark[6] & 5 & 4 & 3 & 2 \\\hline
    \end{tabular}
    \caption{Some characteristics of the previously experimentally reported non-oxidative glycolytic pathways.}
    \footnotetext[1]{count of the total number of reactions used in the pathway: $\sum_e f_e$.}
    \footnotetext[2]{count of the number of unique reactions used in the pathway: $\sum_{e:f_e>0} 1$.}
    \footnotetext[3]{count of the different enzymes used in the pathway.}
    \footnotetext[4]{3-tuple representing the count of the number of reactions using (Xpk, Fpk, Spk) respectively.}
    \footnotetext[5]{count of the number of bimolecular reactions in the pathway.}
    \footnotetext[6]{count of the number of reactions where both the reactant molecules are carbohydrates.}
    \label{tab: pathways_describe}
\end{table}

Reactions involving two carbohydrate molecules as substrates (catalyzed by transaldolase or transketolase enzymes) might be potential kinetic bottlenecks in the pathway. We generate thermodynamic profiles for the four pathways to deduce which reactions might act as thermodynamic bottlenecks. The thermodynamic free-energy profiles indicate that the first reaction catalyzed by the transaldolase in all four pathways is unfavorable. The hydrolysis of fructose-1,6-biphosphate catalyzed by phosphohydrolase has a marginally negative Gibbs free energy difference. This was expected from the free energy differences for the hydrolysis of the five-carbon and seven-carbon biphosphates. Except for the first pathway \ref{tab: bogorad_2a} (which has three thermodynamically unfavorable reactions), all the pathways have only one thermodynamically unfavorable reaction. Increasing the total number of reactions in the pathway did not have an adverse effect on the free energy differences of the individual reactions.

\begin{figure}
    \centering
    \begin{subfigure}{0.49\textwidth}
        \includegraphics[width=\linewidth]{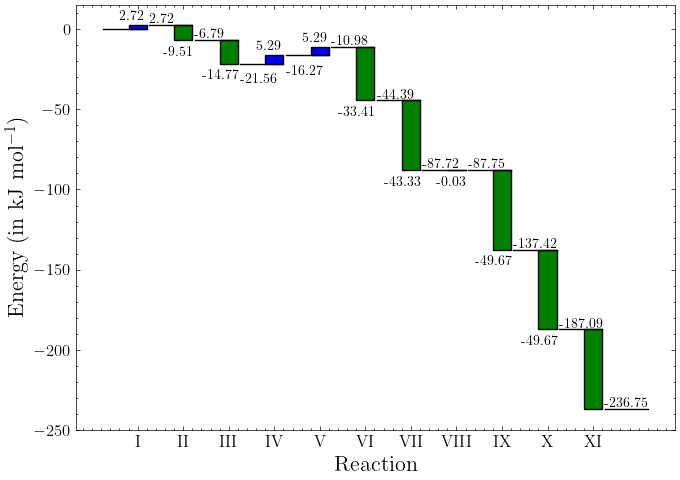}
        \caption{\label{fig: bogorad2a_profile}}
    \end{subfigure}%
    ~
    \begin{subfigure}{0.49\textwidth}
        \includegraphics[width=\linewidth]{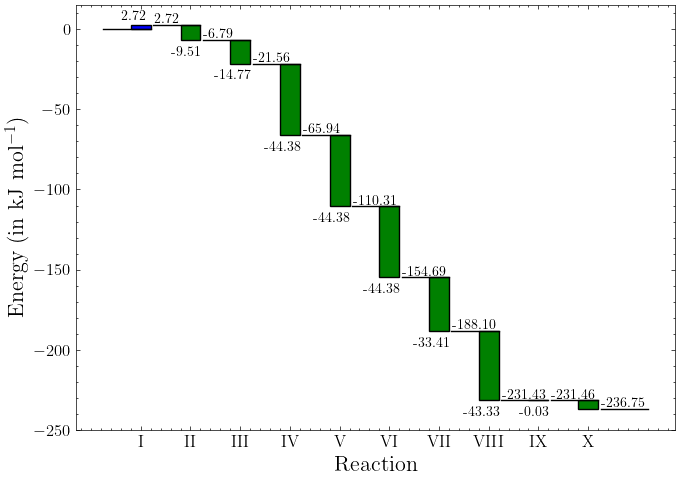}
        \caption{\label{fig: bogorad2b_profile}}
    \end{subfigure}
    \begin{subfigure}{0.49\textwidth}
        \includegraphics[width=\linewidth]{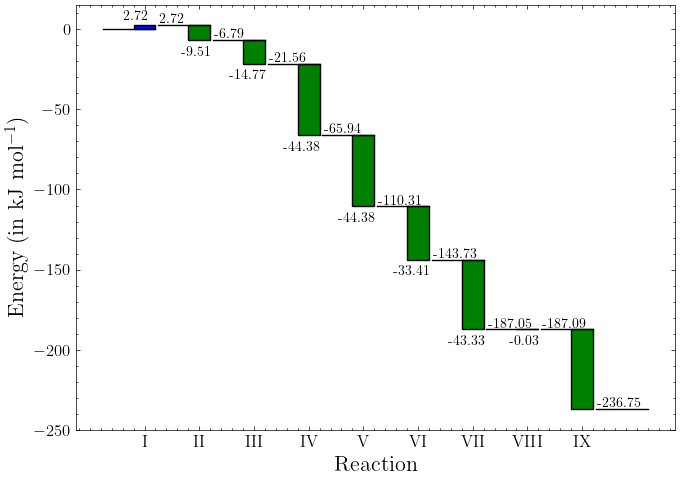}
        \caption{\label{fig: bogorad2c_profile}}
    \end{subfigure}%
    ~
    \begin{subfigure}{0.49\textwidth}
        \includegraphics[width=\linewidth]{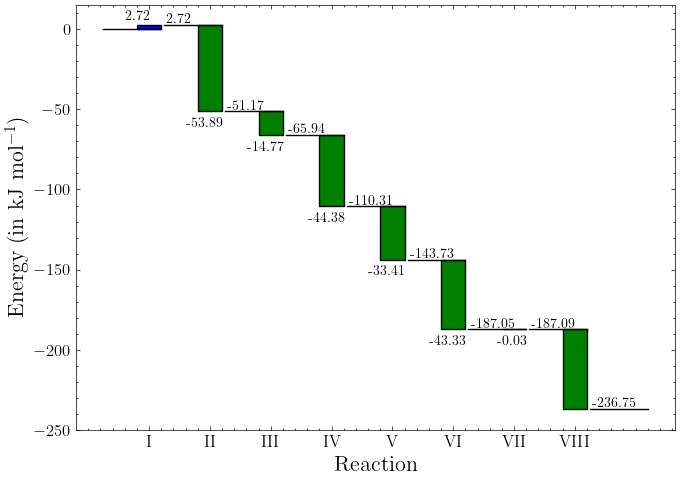}
        \caption{\label{fig: krusemann_profile}}
    \end{subfigure}
    \caption{Thermodynamic free energy profiles for the experimentally reported pathways \ref{tab: bogorad_2a} (top left), \ref{tab: bogorad_2b} (top right), \ref{tab: bogorad_2c} (bottom left) and \ref{tab: krusemann} (bottom right) for non-oxidative glycolysis.}
\end{figure}

Two factors might account for the large number of possible pathways for a non-oxidative glycolytic pathway: the promiscuous activity of phosphoketolase and the biphosphate that might be hydrolyzed. The phosphoketolase acting on different substrates, as shown in reactions \ref{xpk}, \ref{fpk}, and \ref{spk}, have comparable free energy differences: $-44.376$ kJ mol$^{-1}$, $-49.666$ kJ mol$^{-1}$ and $-53.887$ kJ mol$^{-1}$, respectively. Regarding biphosphate hydrolysis, fructose-1,6-biphosphate hydrolysis has a negative Gibbs free energy difference of $0.0322$ kJ mol$^{-1}$, while the hydrolysis of both xylulose-1,6-biphosphate and sedoheptulose-1,7-biphosphate has positive Gibbs free energy differences: $7.386$ kJ mol$^{-1}$ and $0.576$ kJ mol$^{-1}$, respectively. The free energies of reactions can be altered by changing the concentrations of the molecules involved. The hydrolysis of sedoheptulose-1,7-biphosphate is quite close to equilibrium.
\begin{equation*}
    \Delta_r G = -RT \ln \left(\frac{[\text{S7P}][\text{P}_i]}{[\text{SBP}][H_2O]}\right)
\end{equation*}
A concentration ratio of only $1.26$ for [SBP]/[S7P] moves the reaction to thermal equilibrium at $300$ K, considering the unit activities of the water and phosphate molecules. The corresponding value for the ratio of [XBP]/[X5P] so that the reaction is in equilibrium is $19.32$. These concentration gradients can be maintained by the active transport of the molecules involved in the reactions.

\subsection{\label{alternative} Alternative equivalent non-oxidative glycolysis pathways}

\begin{figure}[t]
    \centering
    \includegraphics[width=\linewidth]{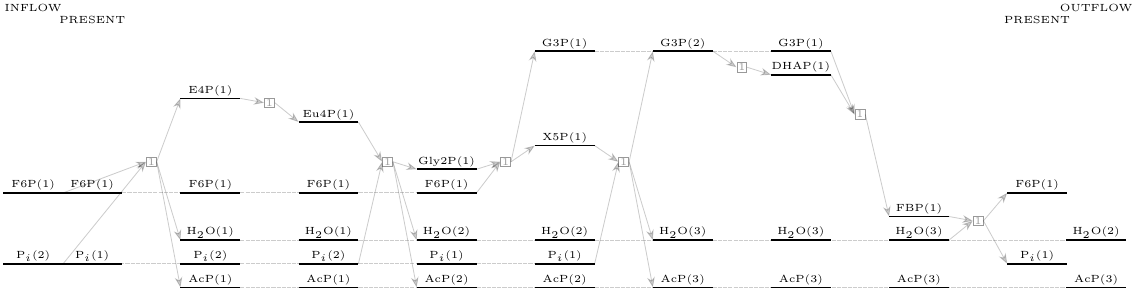}
    \caption{An alternative non-oxidative glycolytic pathway using fructose-1,6-biphosphate and eight reactions.}
    \label{fig: fbp_8_reactions}
\end{figure}

The results of the previous section indicate that the hydrolysis of fructose-1,6-biphosphate is thermodynamically favored compared to immediate higher or lower sugars. An exhaustive search for pathways in the generated molecular space indicates that there is an alternative pathway, shown in Figure \ref{fig: fbp_8_reactions}, using eight different reactions (each used once) with the biphosphate of fructose to achieve the non-oxidative glycolytic pathway.

\begin{figure}[ht]
    \centering
    \begin{subfigure}{0.49\textwidth}
    \centering
    \begin{tiny}
        \begin{tabular}{r r r c r l l l l l}
            F6P &+ &P$_i$ &$\overset{\text{PK}}\longrightarrow$ &&E4P &+ &AcP &+ &H$_2$O \\
            & & E4P &$\overset{\text{AlKe}}\longrightarrow$ &&Eu4P \\
            Eu4P &+ &P$_i$ &$\overset{\text{PK}}\longrightarrow$ &&Gly2P &+ &H$_2$O &+ &AcP \\
            Gly2P &+ &F6P &$\overset{\text{TAL}}\longrightarrow$ &&X5P &+ &G3P \\
            X5P &+ &P$_i$ &$\overset{\text{PK}}\longrightarrow$ &&G3P &+ &AcP &+ &H$_2$O \\
            & &G3P &$\overset{\text{AlKe}}\longrightarrow$ &&DHAP \\
            G3P &+ &DHAP &$\overset{\text{AL}}\longrightarrow$ &&FBP \\
            FBP &+ &H$_2$O &$\overset{\text{PHL}}\longrightarrow$ &&F6P &+ &P$_i$ \\\hline
            F6P &+ &2P$_i$ &$\longrightarrow$ &3&AcP &+ 2&H$_2$O
        \end{tabular}
    \end{tiny}
    \caption{an alternative pathways}
    \label{tab: alt_fbp_8reactions}
    \end{subfigure}%
    ~
    \begin{subfigure}{0.49\textwidth}
        \centering
        \includegraphics[width=0.75\linewidth]{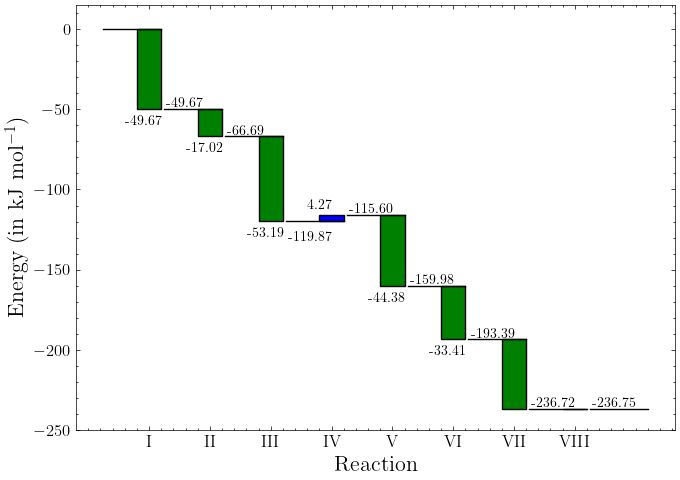}
        \caption{its thermodynamic free energy profile}
        \label{fig: alt_fbp_8reactions_profile}
    \end{subfigure}
    \caption{The sequence of reactions and details of the alternative non-oxidative glycolytic pathway which uses eight reactions and fructose-1,6-biphosphate}
\end{figure}

This pathway uses five enzymes: phosphoketolase, aldose-ketose isomerase, transaldolase, aldolase, and phosphohydrolase, as detailed in Table \ref{tab: alt_fbp_8reactions}. The thermodynamic free energy profile for this pathway is plotted in Figure \ref{fig: alt_fbp_8reactions_profile}. Only one reaction, the bimolecular aldol addition reaction involving two carbohydrate molecules catalyzed by a transaldolase enzyme, is thermodynamically unfavorable ($\Delta_r G = +4.27$ kJ mol$^{-1}$). The experimentally reported pathway with eight reactions using the same biphosphate also has a thermodynamically unfavorable reaction, although one with a smaller positive Gibbs free energy difference ($\Delta_r G = +2.72$ kJ mol$^{-1}$). On the basis of this thermodynamic analysis, one might conjecture that both of these reactions might proceed in parallel in a bioreactor if a phosphoketolase acting on erythrulose is added to the system.

Formulating the search problem as an ILP query allows us to exhaustively search for non-oxidative glycolytic pathways in the expanded molecular network with the queried length. The results of this search are summarized in Table \ref{tab: pathways_length}. The length of a pathway is the number of reactions in the pathway. If a reaction is used twice in the pathway, it is counted twice. There are only two pathways of length seven: both are described in Section \ref{seven}. There are twelve pathways that use eight reactions in total, but seven different reactions, and none of them use fructose-1,6-biphosphate. The shortest pathways that use fructose-1,6-biphosphate are of length eight, using all different reactions. They are described in Sections \ref{experiment} and \ref{alternative}. As the length of the pathway allowed increases, the number of solutions returned by the ILP solver increases, as indicated in the last three rows of Table \ref{tab: pathways_length}. The complete list of the pathways enumerated in the table can be found in the associated GitHub repository. 

\subsection{Other pathways}

Expanding the molecular space using the eight rules resulted in $81$ molecules in the reaction network, of which $39$ were phosphates of carbohydrates, $14$ were biphosphates of carbohydrates, $23$ were carbohydrates, and $5$ were other small molecules. These other small molecules included phosphoric acid, water, ethanoic acid (isomeric to glycolaldehyde), methanal, and 3-hydroxypropionic acid (isomeric to glyceraldehyde). 3-Hydroxypropionic acid was listed as a "proof-of-concept" product of biosynthesis. It is identified as a precursor to prop-2-enoic acid (acrylic acid) \cite{hydroxypropanoate} and the biodegradable polymer poly(3-hydroxypropionic acid), or P(3-HPA) \cite{p3_hpa1, p3_hpa2}. Moreover, acrylic acid can be used to synthesize acrylates, which are monomers for versatile polymers: poly(methyl methacrylate) (PMMA, a lightweight and shatter-resistant alternative to glass) and poly(ethyl cyanoacrylate) (PECA, super glues), to name a few. Both ethanoic acid (acetic acid) and methanal (formaldehyde) have numerous industrial applications in addition to being precursors to polymeric units. Therefore, it might be worthwhile to explore the pathways that lead to these molecules in the network. The pathways producing more than one methanal molecule could not be enumerated in this network. Given that there are multiple alternative more economic reaction sequences that lead to the production of methanal \cite{hcho_synthesis}, its production through this biosynthetic method with such a low carbon efficiency might not be justified.

\begin{figure}[ht]
    \centering
    \begin{subfigure}{0.25\textwidth}
        \includegraphics[width=\linewidth]{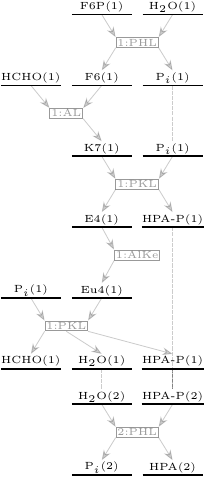}
        \caption{\label{fig: path1}}
    \end{subfigure}%
    \hfill
    \begin{subfigure}{0.25\textwidth}
        \includegraphics[width=\linewidth]{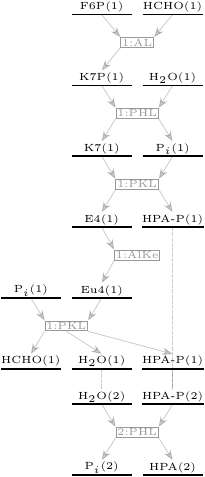}
        \caption{\label{fig: path2}}
    \end{subfigure}%
    \hfill
    \begin{subfigure}{0.25\textwidth}
        \includegraphics[width=\linewidth]{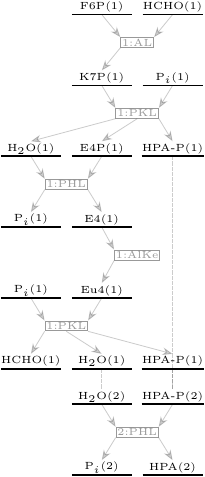}
        \caption{\label{fig: path3}}
    \end{subfigure}
    \begin{subfigure}{0.25\textwidth}
        \includegraphics[width=\linewidth]{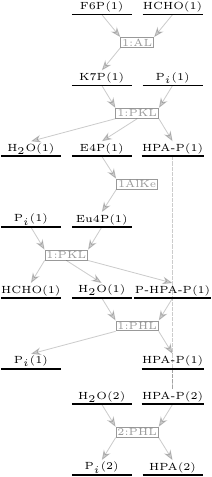}
        \caption{\label{fig: path4}}
    \end{subfigure}%
    \hfill
    \begin{subfigure}{0.25\textwidth}
        \includegraphics[width=\linewidth]{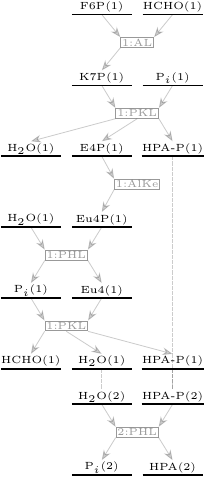}
        \caption{\label{fig: path5}}
    \end{subfigure}%
    \hfill
    \begin{subfigure}{0.25\textwidth}
        \includegraphics[width=\linewidth]{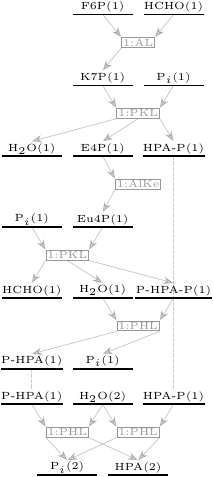}
        \caption{\label{fig: path6}}
    \end{subfigure}
    \caption{The six pathways leading to 3-hydroxypropionic acid implementing the net reaction \ref{hpa}. While the first five pathways \ref{fig: path1}-\ref{fig: path5} use six unique reactions (the last reaction is used twice), the last pathway \ref{fig: path6} uses seven distinct reactions.}
    \label{fig: hpa_paths}
\end{figure}

\subsubsection{Pathways to 3-hydroxypropioic acid}
Since 3-hydroxypropionic acid is a molecule containing three carbon atoms, the optimal efficiency would be achieved when two molecules are produced from one molecule of fructose-6-phosphate, leading to the overall reaction:
\begin{equation}
    \text{F6P } + \text{H}_2\text{O} \longrightarrow 2\text{HPA } + \text{P}_i \text{\hspace{2cm}} \Delta_r G = -114.428 \text{ kJ mol}^{-1}
    \label{hpa}
\end{equation}
In the generated reaction network, \emph{six} such pathways were enumerated that carried out the overall reaction \ref{hpa} in seven steps. Five of these pathways used six unique reactions (one reaction was used twice), while the last pathway used seven distinct reactions, as shown in Figure \ref{fig: hpa_paths}. All pathways require four of the seven enzymes used to generate the network: aldolase, aldose-ketose isomerase, phosphohydrolase, and phosphoketolase.

\begin{figure}[!ht]
    \centering
    \begin{subfigure}{0.32\textwidth}
        \includegraphics[width=\linewidth]{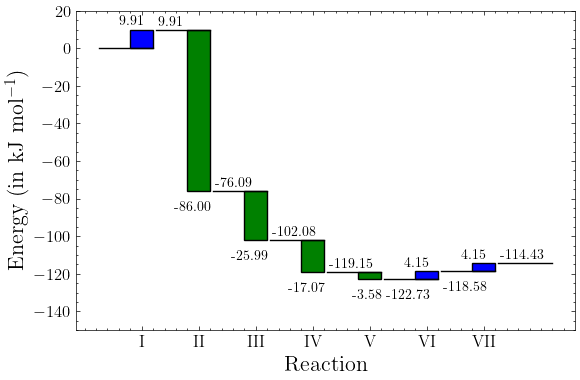}
        \caption{\label{fig: path1_profile}}
    \end{subfigure}%
    ~
    \begin{subfigure}{0.32\textwidth}
        \includegraphics[width=\linewidth]{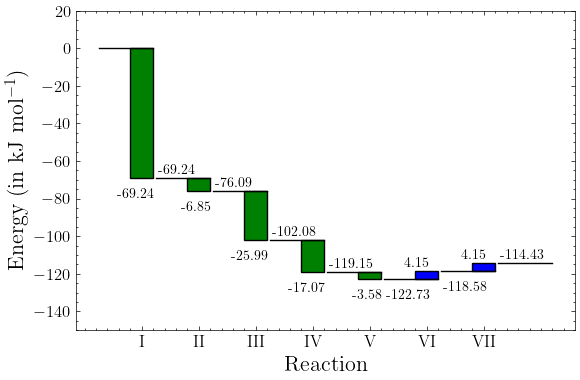}
        \caption{\label{fig: path2_profile}}
    \end{subfigure}%
    ~
    \begin{subfigure}{0.32\textwidth}
        \includegraphics[width=\linewidth]{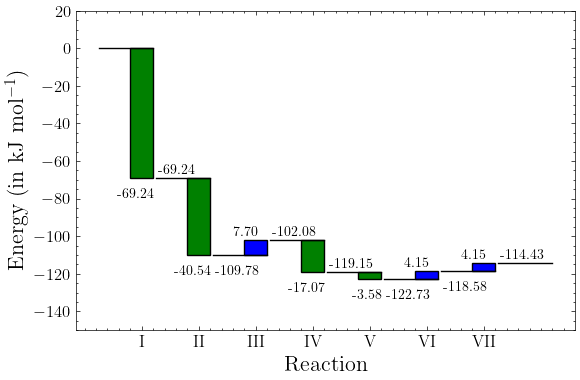}
        \caption{\label{fig: path3_profile}}
    \end{subfigure}
    \begin{subfigure}{0.32\textwidth}
        \includegraphics[width=\linewidth]{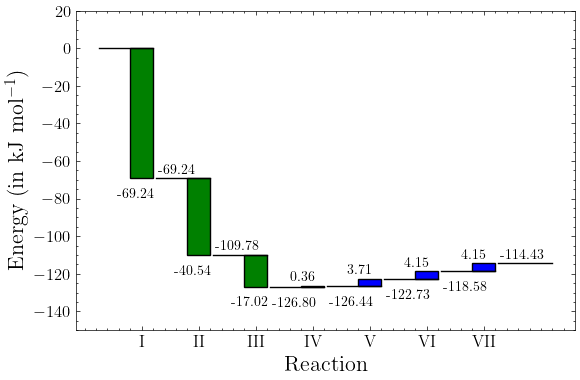}
        \caption{\label{fig: path4_profile}}
    \end{subfigure}%
    ~
    \begin{subfigure}{0.32\textwidth}
        \includegraphics[width=\linewidth]{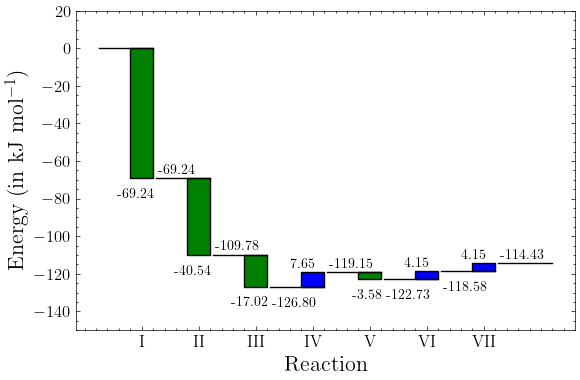}
        \caption{\label{fig: path5_profile}}
    \end{subfigure}%
    ~
    \begin{subfigure}{0.32\textwidth}
        \includegraphics[width=\linewidth]{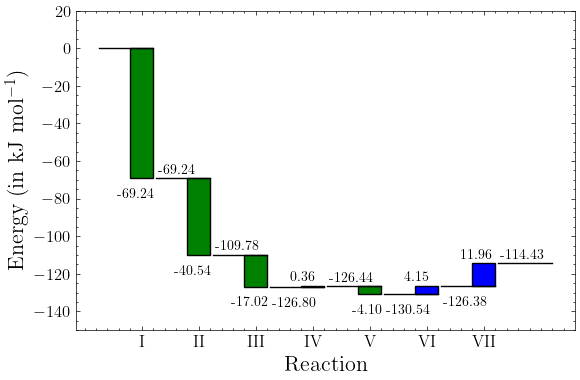}
        \caption{\label{fig: path6_profile}}
    \end{subfigure}
    \caption{Thermodynamic profiles for the six corresponding pathways listed in Figure \ref{fig: hpa_paths} leading to 3-hydroxypropionic acid.}
    \label{fig: hpa_profiles}
\end{figure}

The pathway in \ref{fig: path2_profile} uses only two reactions with a positive $\Delta_r G$ (it is, in fact, the same reaction implemented twice), while all other pathways have more thermodynamically unfavorable reactions. This could be processed separately, where the 3-hydroxypropionic acid phosphate is collected as a product and the phosphate group is removed separately. However, in a reactor where non-oxidative glycolysis is allowed to run in parallel, the net reaction \ref{hpa} would be overshadowed by it because it has a much lower change in free energy.

\subsubsection{Pathways to acetic acid}
Preserving the carbon efficiency would lead to the following overall reaction, with three molecules of acetic acid as the product for every molecule of fructose-3-phosphate: 
\begin{equation}
    \text{F6P } + \text{H}_2\text{O} \longrightarrow 3\text{AcH } + \text{P}_i \text{\hspace{2cm}} \Delta_r G = -286.711 \text{ kJ mol}^{-1}
    \label{acetate}
\end{equation}
Sixty-one pathways were enumerated that affect the overall reaction \ref{acetate} in ten steps. Three of these pathways used six different reactions (one reaction was used three times and a second reaction twice), six pathways used seven different reactions, while the remaining $52$ used eight distinct reactions. In addition, acetylphosphate in hydrolysis also yields acetic acid. Therefore, in theory, any pathway that produces acetylphosphate can also be used to produce acetic acid. Moreover, formaldehyde was also obtained in the expanded molecular space. However, there might be better and more efficient ways to produce formaldehyde than using enzymes. Hence, such pathways are not discussed here.

\section{Conclusion}
In this work, we attempted to generate Gibbs free energy profiles for pathways implementing non-oxidative glycolysis in the expanded molecular space. The expansion of the molecular space using reaction templates for the action of each enzyme, along with the ILP-based search for pathways, allows us to explore the molecular space without bias. Although the use of Gibbs free energy under standard physical conditions might not always mirror what happens inside a cell, it can assist in elucidating which pathways might be favored by thermodynamics and identifying the bottlenecks in each pathway.

A recurring reaction in the pathways with a positive change in Gibbs free energy was the hydrolysis of a phosphate group from a molecule, whether it involved a carbohydrate biphosphate producing a monophosphate or a carbohydrate monophosphate producing a carbohydrate molecule. Because this reaction involves a water molecule, performing it in an aqueous medium might shift the energetics of the reaction toward the dephosphorylated products and phosphoric acid.

Another interesting result from the search query for pathways that implement non-oxidative glycolysis was the involvement of phosphoketolase acting on the phosphate of an erythrose phosphate molecule. If one could develop a variant of the phosphoketolase enzyme that is active on this four-carbon ketose, it might be possible to achieve the pathways returned by the ILP solver, provided that the required concentration gradients are maintained in the system. Expanding the molecular space with generic rules (irrespective of whether the resulting reactions have been previously reported) provides insights into numerous possibilities for pathways implementing non-oxidative glycolysis, a few of which have been studied in greater detail in this work.

Although existing work focuses on pathways ensuring optimal carbon efficiency during non-oxidative glycolysis, there is a tendency to prioritize obtaining acetyl phosphate as the product because of its metabolic significance. However, other molecules in the generated network might also be of interest, one of them being 3-hydroxypropionic acid. Under laboratory conditions, the products from non-oxidative glycolysis are observed because they have lower Gibbs free energy. The larger Gibbs free energy difference makes the product, acetyl phosphate, $e^{\frac{-236.75}{-114.28}} \sim 8$ times more likely to be observed than 3-hydroxypropionic acid. However, pathways to 3-hydroxypropionic acid require only four of the seven enzymes needed for the originally proposed non-oxidative glycolysis. Although non-oxidative glycolysis might still be achieved using six of these seven enzymes, the optimal carbon efficiency cannot be achieved with only the four enzymes needed to produce 3-hydroxypropionic acid. Therefore, in a system with only four of these enzymes, the pathway producing 3-hydroxypropionic acid might be expected to dominate over some subset of non-oxidative glycolysis.

The method of analyzing pathways satisfying the constraints of mass balance through thermodynamic free energies provides insights into pathways that might be promising under practical conditions. This work identifies bottlenecks in the pathways that could require manipulation of the concentrations of the molecules involved in those reactions to make them favorable. Lastly, it highlights potentially useful by-products in the expanded network and offers directions for further laboratory exploration to synthesize the identified compounds. The results discussed in this work can be found in more detail in \cite{data_repository}.

\begin{ack}
The author expresses his gratitude to the Institut for Matematik og Datalogi, Syddansk Universitet, for providing the necessary computational resources for this work.
\end{ack}

\appendix
\section{Appendix: Molecules and graph transformation rules used}\label{secA1}

Table \ref{tab:mol_table} lists the starting molecules used to expand the molecular space using the graph transformation rules depicted in Table \ref{tab:transformation_rules}. These rules represent templates for the reactions effected by the corresponding enzyme. In the rules $R$ represents a generic group that meets the valencies of the formulae that can match any subgraph in the representation of the molecule as a graph. Table \ref{tab:more_mol_table} lists some of the interesting molecules that have been used in the pathways studied in this work.

\begin{table}[h!]
    \resizebox{\textwidth}{!}{%
    \begin{tabular}{r | c c c c c c c c c c}
        aldolase & \chemfig{R'-[:30](=[:90]O)-[:-30]H} & + & \chemfig{H-[:-30](-[:120]H)(-[:-90]OH)-[:30](=[:90]O)-[:-30]R} & $\longrightarrow$ & \chemfig{R'-[:30](-[:90]OH)(-[:-120]H)-[:-30](-[:-90]OH)-[:30](=[:90]O)-[:-30]R} & \\\\
        aldoketose & & & \chemfig{H-[:30](=[:90]O)-[:-30](-[:-90]OH)(-[:60]H)-[:30]R} & $\longrightarrow$ & \chemfig{H-[:30](-[:90]OH)(-[:-120]H)-[:-30](=[:-90]O)-[:30]R} \\\\
        ketoaldose & & & \chemfig{H-[:30](-[:90]OH)(-[:-120]H)-[:-30](=[:-90]O)-[:30]R} & $\longrightarrow$ & \chemfig{H-[:30](=[:90]O)-[:-30](-[:-90]OH)(-[:60]H)-[:30]R} \\\\
        phosphohydrolase & H$_2$O & + & \chemfig{R-[:-30]O-[:30]P(=[:90]O)(-[:-30]OH)(-[:-90]OH)} & $\longrightarrow$ & ROH & + & \chemfig{HO-[:30]P(=[:90]O)(-[:-30]OH)(-[:-90]OH)} \\\\
        phosphoketolase & \chemfig{HO-[:30]P(=[:90]O)(-[:-30]OH)(-[:-90]OH)} & + & \chemfig{R-[:-30](-[:-90]OH)(-[:120]H)-[:30](=[:90]O)-[:-30](-[:-90]OH)(-[:60]H)-[:30]R'} & $\longrightarrow$ & \chemfig{R-[:30](=[:90]O)-[:-30]H} & + & \chemfig{HO-[:30]P(=[:90]O)(-[:-90]OH)-[:-30]O-[:30](=[:90]O)-[:-30](-[:-120]H)(-[:-60]H)-[:30]R'} & + & H$_2$O \\\\
        transaldolase & \chemfig{R'-[:30](=[:90]O)-[:-30]H} & + & \chemfig{HO-[:-30](-[:-120]H)(-[:-60]H)-[:30](=[:90]O)-[:-30](-[:-90]OH)(-[:60]H)-[:30](-[:90]OH)(-[:-60]H)-[:-30]R} & $\longrightarrow$ & \chemfig{R-[:30](=[:90]O)-[:-30]H} & + & \chemfig{HO-[:-30](-[:-120]H)(-[:-60]H)-[:30](=[:90]O)-[:-30](-[:-90]OH)(-[:60]H)-[:30](-[:90]OH)(-[:-60]H)-[:-30]R'} \\\\
        transketolase & \chemfig{R'-[:30](=[:90]O)-[:-30]H} & + & \chemfig{HO-[:-30](-[:-120]H)(-[:-60]H)-[:30](=[:90]O)-[:-30](-[:-90]OH)(-[:60]H)-[:30]R} & $\longrightarrow$ & \chemfig{R-[:30](=[:90]O)-[:-30]H} & + & \chemfig{HO-[:-30](-[:-120]H)(-[:-60]H)-[:30](=[:90]O)-[:-30](-[:-90]OH)(-[:60]H)-[:30]R'} \\\\
    \end{tabular}
    }
    \caption{List of the graph transformation rules used as templates for the enzymatic reactions}
    \label{tab:transformation_rules}
\end{table}

Of the $81$ molecules in the expanded network, $6$ were small molecules, $22$ carbohydrates, $39$ were monophosphates, and $14$ were biphosphates. There were more monophosphates than carbohydrates because a phosphate group could attach to different hydroxyl positions in the carbohydrate molecule, leading to different monophosphate molecules.

\newpage 
\begin{table}[h!]
    \centering
    \resizebox{\textwidth}{!}{%
    \begin{tabular}{rccl}
        \hline
        metabolite name & vertex label & \hspace{1cm} & molecular structure \\
        \hline\\
        water & H$_2$O && H$_2$O \\\\
        phosphoric acid & P$_i$ && \chemfig{P(=[::+90]O)(-[:-30]OH)(-[:-90]OH)(-[:-150]HO)} \\\\
        acetylphosphate & AcP && \chemfig{[:-30]-[:30](=[:90]O)-[:-30]O-[:30]P(=[:90]O)(-[:-90]OH)(-[:-30]OH)} \\\\
        glyceraldehyde 3-phosphate & G3P && \chemfig{(=[:-150]O)-[:-30](-[:-90]OH)-[:30]-[:-30]O-[:30]P(=[:90]O)(-[:-90]OH)(-[:-30]OH)}\\\\
        dihydroxyacetone 3-phosphate & DHAP && \chemfig{(-[:-150]HO)-[:-30](=[:-90]O)-[:30]-[:-30]O-[:30]P(=[:90]O)(-[:-90]OH)(-[:-30]OH)}\\\\
        erythrose 4-phosphate & E4P && \chemfig{(=[:150]O)-[:30](-[:90]OH)-[:-30](-[:-90]OH)-[:30]-[:-30]O-[:30]P(=[:90]O)(-[:-90]OH)(-[:-30]OH)}\\\\
        ribose 5-phosphate & R5P && \chemfig{(=[:-150]O)-[:-30](-[:-90]OH)-[:30](-[:90]OH)-[:-30](-[:-90]OH)-[:30]-[:-30]O-[:30]P(=[:90]O)(-[:-90]OH)(-[:-30]OH)}\\\\
        xylulose 5-phosphate & X5P && \chemfig{(-[:-150]HO)-[:-30](=[:-90]O)-[:30](-[:90]OH)-[:-30](-[:-90]OH)-[:30]-[:-30]O-[:30]P(=[:90]O)(-[:-90]OH)(-[:-30]OH)}\\\\
        fructose 6-phosphate & F6P && \chemfig{(-[:150]HO)-[:30](=[:90]O)-[:-30](-[:-90]OH)-[:30](-[:90]OH)-[:-30](-[:-90]OH)-[:30]-[:-30]O-[:30]P(=[:90]O)(-[:-90]OH)(-[:-30]OH)}\\\\
        sedoheptulose 7-phosphate & S7P && \chemfig{(-[:-150]HO)-[:-30](=[:-90]O)-[:30](-[:90]OH)-[:-30](-[:-90]OH)-[:30](-[:90]OH)-[:-30](-[:-90]OH)-[:30]-[:-30]O-[:30]P(=[:90]O)(-[:-90]OH)(-[:-30]OH)}\\\\
        fructose 1,6-biphosphate & FBP && \chemfig{HO-[:-30]P(=[:-90]O)(-[:90]OH)-[:30]O-[:-30]-[:30](=[:90]O)-[:-30](-[:-90]OH)-[:30](-[:90]OH)-[:-30](-[:-90]OH)-[:30](-[:-30]O-[:30]P(=[:90]O)(-[:-90]OH)(-[:-30]OH))}\\
    \end{tabular}
    }
    \caption{List of the starting molecules used to expand the reaction network}
    \label{tab:mol_table}
\end{table}

\newpage
\begin{table}[h!]
    \centering
    \resizebox{\textwidth}{!}{%
    \begin{tabular}{rccl}
        \hline
        metabolite name & vertex label & \hspace{1cm} & molecular structure \\
        \hline\\
        glycolaldehyde 2-phosphate & Gly2P && \chemfig{(=[:150]O)-[:30]-[:-30]O-[:30]P(=[:90]O)(-[:-90]OH)(-[:-30]OH)} \\\\
        erythrulose 4-phosphate & Eu4P && \chemfig{(-[:150]HO)-[:30](=[:90]O)-[:-30](-[:-90]OH)-[:30]-[:-30]O-[:30]P(=[:90]O)(-[:-90]OH)(-[:-30]OH)}\\\\
        xylulose 1,5-biphosphate & XBP && \chemfig{HO-[:30]P(=[:90]O)(-[:-90]OH)-[:-30]O-[:30]-[:-30](=[:-90]O)-[:30](-[:90]OH)-[:-30](-[:-90]OH)-[:30]-[:-30]O-[:30]P(=[:90]O)(-[:-90]OH)(-[:-30]OH)}\\\\    
        sedoheptulose 1,7-phosphate & SBP && \chemfig{HO-[:30]P(=[:90]O)(-[:-90]OH)-[:-30]O-[:30]-[:-30](=[:-90]O)-[:30](-[:90]OH)-[:-30](-[:-90]OH)-[:30](-[:90]OH)-[:-30](-[:-90]OH)-[:30]-[:-30]O-[:30]P(=[:90]O)(-[:-90]OH)(-[:-30]OH)}\\\\
        fructose 1,6-biphosphate & FBP && \chemfig{HO-[:-30]P(=[:-90]O)(-[:90]OH)-[:30]O-[:-30]-[:30](=[:90]O)-[:-30](-[:-90]OH)-[:30](-[:90]OH)-[:-30](-[:-90]OH)-[:30](-[:-30]O-[:30]P(=[:90]O)(-[:-90]OH)(-[:-30]OH))}\\\\
        phosphate of a ketoheptose & K7P && \chemfig{HO-[:30]P(=[:90]O)(-[:-90]OH)-[:-30]O-[:30]-[:-30](-[:-90]OH)-[:30](-[:90]OH)-[:-30](-[:-90]OH)-[:30](=[:90]O)-[:-30](-[:-90]OH)-[:30]-[:-30]OH}\\\\
        3-hydroxypropionic acid & HPA && \chemfig{HO-[:30](=[:90]O)-[:-30]-[:30]-[:-30]OH}\\\\
        3-(phosphonooxy)propanoic acid & HPA-P && \chemfig{HO-[:30](=[:90]O)-[:-30]-[:30]-[:-30]O-[:30]P(=[:90]O)(-[:-90]OH)(-[:-30]OH)}\\\\
        3-hydroxypropionic phospahate & P-HPA && \chemfig{HO-[:30]P(=[:90]O)(-[:-90]OH)-[:-30]O-[:30](=[:90]O)-[:-30]-[:30]-[:-30]OH}\\\\
    \end{tabular}
    }
    \caption{List of some molecules encountered in pathways from the reaction network}
    \label{tab:more_mol_table}
\end{table}

\begin{small}
\bibliography{bibliography}
\end{small}

\end{document}